\documentstyle[12pt,epsf]{article}

\newcommand{\be}{\begin{equation}}
\newcommand{\ee}{\end{equation}}
\newcommand{\dd}{{\rm d}}

\begin{document}

\title{New Evidence for  Super-Roughening in Crystalline 
Surfaces with a Disordered Substrate}

\author{Barbara Coluzzi$^{(a)}$, 
Enzo Marinari$^{(b)}$ and Juan J. Ruiz-Lorenzo$^{(a)}$\\[0.5em]
$^{(a)}$  {\small  Dipartimento di Fisica and Infn, Universit\`a di Roma}
   {\small {\em La Sapienza} }\\
{\small   \ \  P. A. Moro 2, 00185 Roma (Italy)}\\[0.3em]
{\small   \tt coluzzi@roma1.infn.it  ruiz@chimera.roma1.infn.it}\\[0.5em]
$^{(b)}$  {\small  Dipartimento di Fisica and Infn, 
Universit\`a di Cagliari}\\
{\small   \ \  Via Ospedale 72, 07100 Cagliari (Italy)}\\[0.3em]
{\small   \tt marinari@ca.infn.it}\\[0.5em]}

\date{January 11, 1997}

\maketitle

\begin{abstract}
We study the behavior of the Binder cumulant related to long distance 
correlation functions of the discrete Gaussian model of disordered 
substrate crystalline surfaces.  We exhibit numerical evidence that 
the non-Gaussian behavior in the low-$T$ region persists on large 
length scales, in agreement with the broken phase being super-rough.
\end{abstract}  

\thispagestyle{empty}
\newpage

\section{\protect\label{S_INT}Introduction}

The disordered discrete Gaussian model ({\sc ddg}) is related to the
$2D$ random phase Sine-Gordon model ({\sc rpsg}). The two models
belong to the same universality class, which includes different
physical systems: for example it is supposed to describe
crystalline surfaces growing upon a disordered
substrate \cite{ton}, as well as $2D$ randomly pinned arrays of flux
lines with the magnetic field parallel to the superconducting plane
\cite{hwa}. 

The standard approach used to investigate this model is 
renormalization group (RG) \cite{ton}-\cite{tsai}.  The application of 
the Mezard-Parisi variational approximation \cite{MEZPAR} (originally 
developed for models with continuous replica symmetry breaking) leads 
to a one-step replica broken solution \cite{gia}, and because of that 
it is problematic \cite{mar}: still the main features found in the 
variational approach are very different from the ones found by using 
the standard RG. The most evident difference is maybe in the behavior 
one expects in the broken phase for the height-height correlation 
function.

A considerable amount of work (see, for instance, 
\cite{gia}-\cite{ll}) has been devoted to the subject.  Contrary to 
some former claims the situation has been shown to be in agreement 
with the existence of a broken super-rough phase, as implied by the RG 
approach, by the numerical simulations of \cite{mar,ll}.

Recently very convincing further evidence for the system being 
super-rough down to $T=0$ has been established in 
references \cite{blas}-\cite{shapir_last}, by exact computations of the 
ground state of finite volume samples of the {\sc ddg}.

In this letter we focus on the non-Gaussian behavior of the {\sc ddg} in 
the broken phase.  We show that it persists on large length scales.
This is definitely not compatible with the predictions of the 
variational theory\footnote{We remind again the reader that we are 
here in the framework of a one step replica broken solution, and that 
this kind of criticisms \cite{mar} does not apply to the variational 
theory when describing a continuum breaking \cite{MEZPAR}.}.
On the contrary our data are well compatible with the behavior 
suggested by RG computations.

Labeling with $ \{ d_i \} $ the integer valued dynamical variables
and by $ \{ \eta_i \} $ the quenched disorder, the Hamiltonian of the {\sc
ddg} model is

\begin{equation}
{\cal H}[\phi] \equiv \frac{\kappa}{2} \sum_{<ij>} 
( \phi_i-\phi_j )^2 ,\hspace{1.5cm}
\phi_i \equiv d_i+ \eta_i  \ ,
\end{equation}
where the sum runs over first neighboring sites of a bidimensional lattice.
The model is related to the limit of coupling constant $\lambda
\rightarrow \infty$ of the {\sc rpsg}

\begin{equation}
{\cal H}[ \phi] \equiv \frac{\kappa}{2} \sum_{<ij>} 
\left ( \phi_i-\phi_j \right )^2 - \lambda \sum_i \cos 
\left ( 2 \pi (\phi_i - \eta_i) \right ) \ ,
\protect\label{E-HDISCR}
\end{equation}
where now ${ \phi_i}$ are the continuous dynamical variables.

The partition function is defined as

\be
  Z_{\eta}\equiv \sum e^{-\beta {\cal H}}\ ,
\ee
where $\beta=\frac{1}{T}$ is the inverse temperature of the problem.

A relevant observable quantity is the correlation function
defined by

\begin{equation}
C(r,T) \equiv \overline{ \langle ( \phi ( {\bf r_0} ) - 
\phi ( {\bf r_0} + {\bf r} ))^2 \rangle }\ .
\end{equation}
In the high-$T$ phase thermal fluctuations make the quenched disorder
irrelevant. Both renormalization group and the variational theory
predict a Gaussian behavior with a logarithmic growth of the
height-height correlation function

\begin{equation}
C_{T > T_c}(r,T) \simeq \frac{T}{k \pi} \log r \ .
\end{equation}
The critical temperature is expected in both approaches to be 
$T_c= \frac{\kappa}{ \pi}$.

The renormalization group approach find that for $T<T_c$ one has
a super-rough broken phase  characterized by

\begin{equation}
C^{\rm RG}_{T<T_c}(r,T) \simeq a \log r + b \log^2 r \ ,\hspace{1.5cm}
b=\frac{2}{\pi^2} \left ( \frac{T_c-T}{T_c} \right)^2 \ ,
\end{equation}
where $a$ is a non-universal coefficient.

On the other hand the Gaussian Ansatz of the variational
approximation hints for no $\log^2r$ contribution, and the broken phase 
turns out to be
described by a one-step replica symmetry broken solution with

\begin{equation}
C^{\rm VAR}_{T<T_c}(r,T) \simeq \frac{T_c}{k \pi} \log r \ ,
\end{equation}
i.e. where the slope of the logarithmic term freezes at the critical point
$T_c$.

Let us recall in a few lines some of the main numerical results 
relevant for the problem.  The authors of \cite{bat} studied {\sc 
rpsg} without being able to detect any signature of the transition 
when measuring static quantities, probably \cite{rie} because of the 
small $\lambda$ value they used.  Indeed for such small values of the 
coupling the difference from the pure case becomes sizable only on 
very large length scales.  In \cite{cush} numerical estimates for the 
correlation function $C(r)$ of the {\sc ddg} model were found to be 
compatible with the picture expected from the variational theory.  
Finally, evidence for the $(\log r)^{2}$ contribution to $C(r)$ in the 
broken phase has been obtained in the case of the {\sc ddg} model 
\cite{mar} and in that of {\sc rpsg} for different $\lambda$ values 
\cite{ll}.

\section{\protect\label{S_ARES}The Binder Parameter}

In this note we will mainly discuss about the distribution function of 
the height-height correlation functions at distance $r$. We define the 
probability distribution 

\begin{equation}
{\cal P} [ \Delta(r),T ] \equiv 
\overline{ \langle  \delta \left [ \Delta ({\bf r}) - ( \phi ( {\bf r_0} ) - 
\phi ( {\bf r_0} + {\bf r} ) )   \right ] \rangle }\ ,
\protect\label{E-PROBAB}
\end{equation}
where by $\langle\cdot\cdot\cdot\rangle $ we denote the thermal 
average (and here also an average over different values of ${\bf r_0}$), and 
by $\overline{\cdot\cdot\cdot}$ we denote the average over disorder.  
For sake of computational simplicity in the following we will only 
consider displacements ${\bf r}$ of the form $(r,0)$ 
or $(0,r)$.  The 
second moment of ${\cal P}$ is the usual height-height correlation 
function $C({\bf r})$, the focus of the investigation of \cite{mar,ll}.  
Here we will try to use the knowledge of the full probability 
distribution (\ref{E-PROBAB}) in order to gather more information 
about the system.

To characterize the probability distribution we measure

\begin{equation}
D(r,T) \equiv 3 \left( \overline{ \langle \Delta^2 (r) \rangle } \right)^2 - 
\overline{ \langle \Delta^4 (r) \rangle}\ ,
\end{equation}
that together with $C(r)$ allows us to define
the Binder cumulant of ${ \cal P}$:

\begin{equation}
B(r,T) \equiv \frac{1}{2} \left ( 3 - \frac{ \overline{ \langle \Delta^4 (r) 
\rangle } }
{ ( \overline{ \langle \Delta^2 (r) \rangle } )^2 } \right ) \ .
\end{equation}
In the thermodynamic limit a value $B=0$ characterizes a Gaussian 
behavior. 

The work of \cite{mar} was based on the analysis of the Binder 
parameter for $r=1$, that was providing evidence for a non Gaussian 
behavior in the low-$T$ region.  We will try here to answer some 
questions that are still open after \cite{mar}, looking at the long 
distance behavior of correlation functions: that will allow us to 
exhibit more evidence for a clear non-gaussian behavior. A purely 
Gaussian behavior could indeed be hidden for small $r$ by short 
distance effects, and manifest itself only in the large $r$ region. 
Analyzing $B(r)$ in the large $r$ region would make this effect clear.

In order to understand what to expect for $B$ we will use 
renormalization group (replica symmetric, to start with).  We will 
mainly follow the approach described by Bernard in his Les Houches 
lecture notes \cite{ber}. In the field theoretical RG approach we 
start from the continuum version of (\ref{E-HDISCR}) by writing

\be
  S = \int \frac{\dd^{2}x}{4\pi}
  \left(
    \frac{\kappa}{2}
    \left(
      \partial_\mu\Phi(x)
    \right)^2
    -\lambda \cos
    \left(
      \Phi(x)-d(x)
    \right)
  \right)\ ,
  \protect\label{E-HCONTI}
\ee
where the $\Phi(x)$ are the basic fields of the theory, and the $d(x)$ 
are the quenched random field which make the system disordered.  
Universality is used to argue that these different systems exhibit 
the same critical behavior.  Following Bernard renormalization calls 
for a generalization of this model: one introduces, in addition to the 
random phases, a random potential.  If one would not do that at the 
start the random potential would in any case be generated by 
renormalization. So one writes

\be
  S = \int \frac{\dd^{2}x}{4\pi}
  \left(
    \frac{\kappa}{2}
    \left(
      \partial_\mu\Phi(x)
    \right)^2
    - A_\mu(x) \partial_\mu \Phi(x)
    -\xi(x)e^{i\Phi(x)}
    -\xi^{*}(x)e^{-i\Phi(x)}
  \right)\ ,
  \protect\label{E-HMODIF}
\ee
where the quenched fields $\xi$ are distributed according to

\be
  P[\xi] = e^{-\frac{1}{2\sigma}\int 
  \frac{\dd^{2}x}{4\pi}\xi\xi^{*}}\ ,
\ee
and the field $A_\mu$ can be written, by noticing that the rotational 
part decouples, as $A_\mu(x)\equiv\partial_\mu\Lambda(x)$, and it is 
distributed as

\be
  P[A_\mu] = e^{-\frac{1}{2g}\int 
  \frac{\dd^{2}x}{4\pi}\left(\partial_\mu\Lambda\right)^2}\ .
\ee
A $U(1)$ symmetry guarantees to the model some remarkable properties. 
For example the $g$-dependence of the correlation functions of the 
vertex operator $e^{i\alpha\Phi}$ can be factorized. 

Now one has to play the usual replica trick (at this stage with exact 
replica symmetry by definition). One writes an effective action for 
$n$ replica's of the system, and takes the $n\to 0$ limit. The 
$\beta$ functions of the model (for the running of $\kappa$, $g$ and 
$\sigma$) turn out to be

\begin{eqnarray}
    \nonumber
	\beta_\sigma & = & 2x\sigma-2\sigma^2+ \ldots\ ,
	\label{A-BETUNO}  \\
    \nonumber
	\beta_g & = & \frac{\sigma^2}{2} + \ldots \ ,
	\label{A-BETDUE}  \\
	\beta_\kappa & = & 0\ ,
	\label{A-BETTRE}
\end{eqnarray}
where we have defined $x\equiv\frac{\kappa-\kappa_c}{\kappa}$, and 
$\kappa$ is not renormalized. So in the low $T$ 
phase, for $\kappa>\kappa_c$, it exists a non trivial infrared fixed 
point at $\sigma_*$, with

\be
  \sigma_* = x + \ldots \mbox{ for } x \ll 1\ .
\ee
At $\sigma_*$ we have $\beta_g=\frac{x^2}{2}$, i.e. $g$ still flows 
(see Bernard \cite{ber} for the characterization of such a {\em run 
away fixed point}).

We are interested in RG predictions for correlation functions. At the 
infrared fixed point one finds that

\begin{equation}
  G(r) 
  \equiv 
  \overline{ \langle 
  e^{ i \alpha ( \phi ( {\bf r} ) - \phi ({\bf 0})) } 
  \rangle_{*} }
  = r^{-2\gamma_{*}} 
  e^{ -\frac{ \alpha^2  \beta_{g*}}{2 \kappa^2} (\log r)^2} \ ,
\end{equation}
where $\gamma_*$ is the anomalous dimension at the fixed point, 
$\gamma_*=\frac{\alpha^2}{\kappa}\rho_*+O(\alpha^4)$, and 
$\rho_*=1+O(x)$.  We have already noticed that $\beta_{g*} = 
\frac{x^2}{2}$ (and it gives the large distance $(\log r)^{2}$ 
behavior of the correlation function).

By expanding $G$ in powers of $\alpha$ one finds that at all orders in 
perturbation theory (in the replica symmetric renormalization group 
approach)

\begin{equation}
\overline{ \langle (\Phi(r) - \Phi(0))^2 \rangle }=
\frac{4 \rho_*}{\kappa} \log r + \frac{\beta_{g*}}{\kappa^2} 
(\log r)^2 \ ,
\end{equation}
and for small $x$

\begin{equation}
\overline{ \langle (\Phi(r) - \Phi(0))^2 \rangle }=
\frac{4}{\kappa} \log r + \frac{x^2}{2\kappa^2} 
(\log r)^2 \ .
\end{equation}

In the same way for the  four point correlation function we find that

\begin{equation}
  \overline{ \langle ( \Phi (  r ) - \Phi (  0 ))^4 \rangle }=
  3 \left ( 
  \frac{4 \rho_*}{\kappa} \log r + \frac{\beta_{*g}}{\kappa^2}(\log r)^2 
  \right )^2 
  - 48 d_* \log r \ ,
\end{equation}
where we have defined $d_*$ the unknown coefficient of the $\alpha^4$ 
contribution to the anomalous dimension $\gamma_*$, which we expect to 
depend from the temperature and that could even be zero.
That means that in the RG approach we find that

\begin{equation}
  D^{\rm RG}_{T<T_c}(r,T) = 48 d_* \log r \ ,
\label{drg}
\end{equation}
where $D^{\rm RG}_{T<T_c}$ does not depend on $\beta^{*}_g$, and

\begin{equation}
  B^{\rm RG}_{T<T_c}(r,T) = \frac {24 d_* \log r}{ 
  ( \frac{4 \rho_*}{\kappa} \log r + 
  \frac{ \beta_{g*}}{\kappa^2}(\log r)^2  )^2} 
  \simeq_{r \gg 1} \frac{24 d_* \kappa^2}
  {\beta^{*}_g \log^3 r}\  .
  \label{brg}
\end{equation}
So, if $d_*\ne 0$, $D(r)$ grows logarithmically with $r$.  On the other 
hand according to the renormalization group picture $\lim_{r 
\rightarrow \infty} B(r) = 0 \simeq (\log r)^{-3}$. These are the 
theoretical predictions we use to interpret our numerical findings.

As a last remark we want to notice that by studying a large $N$ 
version of the Random Phase Sine Gordon model,
Bernard and Bauer \cite{bb} 
found $\beta_{g*}$ to be $O\left(\frac{1}{N^3}\right)$.  This means 
that there is no $(\log r)^2$ contribution to the height-height 
correlation function in the $N \rightarrow \infty$ limit in which the 
Gaussian Ansatz of the variational approach, corresponding to the 
leading order in a $\frac{1}{N}$ expansion, is expected to be exact.

\section{\protect\label{S_NRES}Numerical Results}

We have obtained our numerical data from simulations done on the 
Ape-100 computer \cite{ape}.  We have used square lattices of linear 
size $L=64$ and $L=128$, with periodic boundary conditions.  We have 
fixed the surface tension $\kappa$ to $2$.  We have chosen the 
quenched random variables $\{ \eta_i \}$ uniformly in the range $( - 
\frac{1}{2}, + \frac{1}{2} ]$.

We have simulated in parallel a total of $256$ different realizations 
of the quenched substrate, and two uncoupled replicas for each sample.  
We have used a simple Monte Carlo local dynamics, by proposing to 
update in turn the $\{d_i\}$ by an increment of $\pm 1$.  We have used 
an annealing scheme, in which we have visited in turn decreasing 
values of the temperature $T$ ($T=1.0$, $0.95$, $0.90$, $\dots$, 
$0.40$ for $L=64$ and $T=0.90$, $0.80$, $0.70$, $0.65$, $0.60$, 
$0.45$, $0.35$ for $L=128$).  At each $T$ values the thermalization 
sweeps were $0.5$ million for $L=64$ and $0.7$ million for $L=128$ 
(for further details see \cite{mar}).

We show in figure (1) $B(r)$ as a function of $T$ for different $r \ge 
4$ values.  We plot the data obtained on the smaller lattice, $L=64$, 
since here we had a larger number of $T$ values, but the behavior at 
$L=128$ is very similar.  When comparing $L=64$ and $L=128$ in the 
statistical precision of our runs there is no size dependence for 
$r\le 20$.

\begin{figure}[htbp]
  \begin{center}
    \leavevmode
    \epsfysize=250pt\epsffile{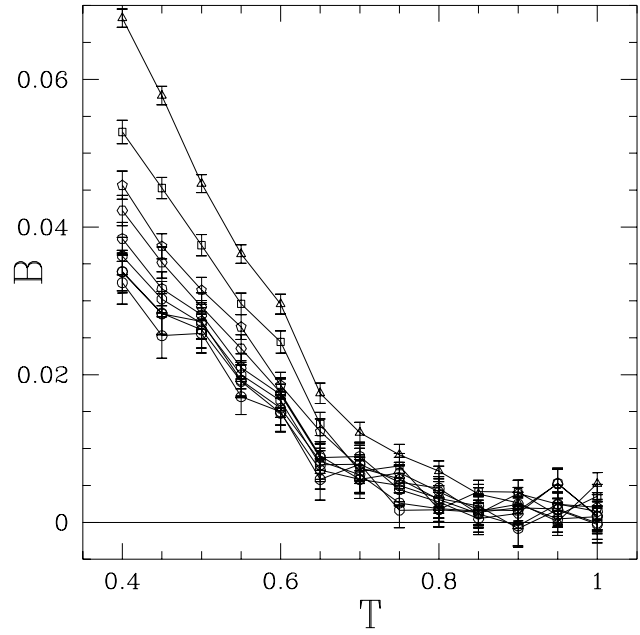}
  \end{center}
  \protect\caption[0]{The Binder parameter as a 
    function of $T$ at different $r \ge 4$ 
    values. $L=64$. Lines are only meant to join neighboring points. Triangles 
    ($r=4$), squares ($r=5$), pentagons ($r=6$), hexagons ($r=7$), etc.}
  \protect\label{fig:1}
\end{figure}

Deviations of $B(r)$ from the Gaussian behavior at small distance 
become evident even for $T \le 0.85$.  The breakdown of the curves 
(for higher $r$ values) in figure (1) is on the contrary compatible with 
the theoretical prediction $T_c=\frac{2}{\pi}$. When looking at 
finite distances one finds a crossover (that does not correspond to a 
true critical behavior) for $T>T_c$ \cite{ll}. Only measuring on 
large lattices real long distance properties one recovers the correct 
critical point, which turns out to coincide with good precision with 
the theoretical predictions.

So, at lower $T$ values the Binder parameter stays non-zero on larger 
length scales. What is even more important is the breaking of the 
slope of $B$ versus $T$ (see figure (1)). From our data we can say 
that for $T<0.65$ the system is surely in its broken phase.

In figure (2) we plot $B(r)$ as a function of $r$ for $T=0.45$ and
$L=128$. Here the fast decay at short distance and the slow decay for 
large $r$ is very clear.

\begin{figure}[htbp]
  \begin{center}
    \leavevmode
    \epsfysize=250pt\epsffile{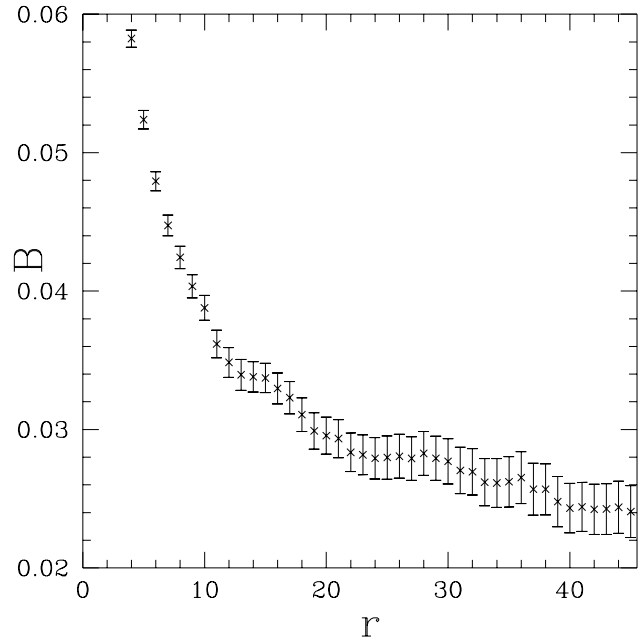}
  \end{center}
  \protect\caption[0]{$B$ as a function of $r$ for $T=0.45$ and $L=128$.}
  \protect\label{fig:2}
\end{figure}

The data shown in figure (2) are qualitatively in very good agreement 
with the prediction of (\ref{brg}), with a non zero value of $d_*$. 

\begin{figure}[htbp]
  \begin{center}
    \leavevmode
    \epsfysize=250pt\epsffile{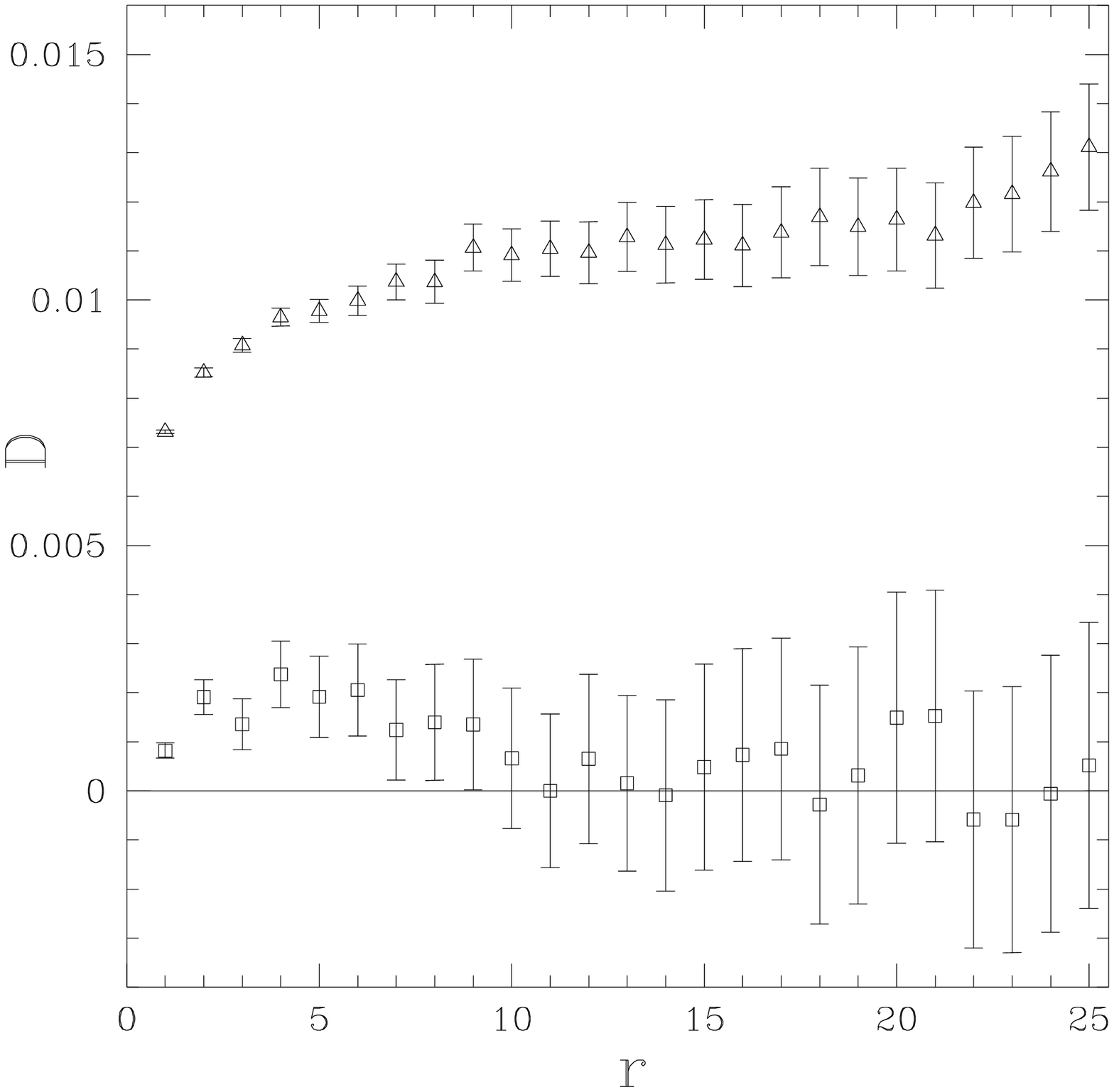}
  \end{center}
  \protect\caption[0]{$D$ as a function of $r$ at $T=0.40$ 
     ($\triangle$) and $T=1.0$ ($\Box$) for $L=64$.} 
  \protect\label{fig:3}
\end{figure}

We show in figure (3) $D(r)$ as a function of $r$ for $L=64$ at the 
highest temperature we have considered ($T=1.0$) and at the lowest one 
($T=0.4$).  That makes clear the difference between the high-$T$ 
region, where $D(r)$ is extremely small even at short distances and 
when increasing $r$ becomes soon compatible with zero, and the broken 
phase where it is definitely non-zero and shows an evident 
increasing behavior.

\begin{figure}[htbp]
  \begin{center}
    \leavevmode
    \epsfysize=250pt\epsffile{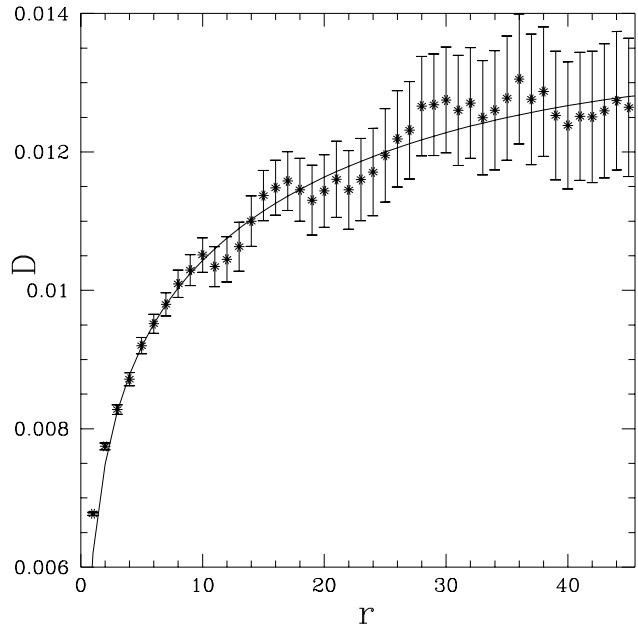}
  \end{center}
  \protect\caption[0]{$D$ as a function of $r$ for $T=0.45$, $L=128$. 
  The line is our best fit to the behavior $D(r)= c_1 P(r)+ c_2$ by using 
  data with $3 \leq r \leq 45$.}
  \protect\label{fig:4}
\end{figure}
  
The fact that in the broken phase $D$ increases with $r$ is clear from 
figure (4), where $D(r)$ is plotted as a function of $r$ for $T=0.45$,
$L=128$.  That clearly shows that the non-Gaussian behavior 
of the model in the broken phase is not a short-distances effect.

We have tried a quantitative analysis of the behavior of $D(r)$, 
at a temperature
well below the critical point. Following \cite{mar,ll}, we use in the
fit the lattice Gaussian propagator

\begin{equation}
P(r) \equiv \frac{1}{2L^2}\sum_{n_1=1}^{L-1} \sum_{n_2=0}^{L-1}
\frac{1-\cos \left( \frac{2 \pi r n_1}{L} \right )}
{2-\cos \left( \frac{2 \pi n_1}{L} \right )-
\cos \left( \frac{2 \pi n_2}{L} \right )} \simeq_{L \gg 1} 
\frac{1}{2 \pi} \left [ \log r + \gamma \log ( 2 \sqrt{2} ) \right ]\  ,
\end{equation}
which enables us to keep finite size effects under control.  The data 
are very well fitted by the expected behavior $D= c_1 P(r)+c_2$.  
Unfortunately errors grow quickly with $r$ and data for $r\ge 40$ 
basically do not influence the fit.

In figure (4) we show our best fit obtained by using data with $ 3 \leq 
r \leq 45$ (disregarding ten more points at short distance does not 
change the results):

\begin{equation}
  10^4 D=(114 \pm 3)P(r) + (36 \pm 3) \ ,
\end{equation}
where errors have been evaluated by using the jack-knife method.  The 
value of the residual $\chi^{2}$ (per degree of freedom) is very good, 
close to $0.2$ (but since the data points are very correlated the 
number does not have necessarily a deep meaning).  The agreement with 
the renormalization group prediction (see equation (\ref{drg})) is 
very good.  Our best numerical estimate for $d_*$ is $\pi c_1 /24$.

\section{\protect\label{S_CON}Conclusions}

Our main conclusion is that the discrete Gaussian model for surfaces
with a disordered substrate in the low-$T$ region is non-Gaussian on
large length scales. Such an evidence was needed to exclude the 
possibility of a short distance effect that could disappear in the 
asymptotic long distance regime.

The picture which emerges from our analysis is therefore incompatible
with the Gaussian variational Ansatz, calling for the broken phase
being super-rough. Finally, our data give numerical evidence for $d_*$
being non-zero, the behavior of $D(r)$ at low temperatures resulting
in good agreement with the logarithmic growth expected from the
renormalization group approach.

\section*{\protect\label{S_ACKNOWLEDGES}Acknowledgments}

We acknowledge useful discussions with Heiko Rieger, David Lancaster 
and Giorgio Parisi.  J. J. Ruiz-Lorenzo is supported by an EC 
HMC(ERBFMBICT950429) grant.

\end{document}